\documentclass[12pt,preprint]{aastex}

\def\kms{km\,s$^{-1}$}

\def\gtrsim{\mathrel{\hbox{\rlap{\hbox{\lower4pt\hbox{$\sim$}}}\hbox{$>$}}}}
\def\lesssim{\mathrel{\hbox{\rlap{\hbox{\lower4pt\hbox{$\sim$}}}\hbox{$<$}}}}

\shorttitle{Progenitor of SN~2009kr}
\shortauthors{Elias-Rosa et al.}

\begin{document}


\title{The Massive Progenitor of the Type II-Linear Supernova
  2009kr\footnote{Based in part on observations made with the NASA/ESA
    {\it Hubble Space Telescope (HST)}, obtained from the Data Archive
    at the Space Telescope Science Institute, which is operated by the
    Association of Universities for Research in Astronomy (AURA),
    Inc., under NASA contract NAS 05-26555; the 6.5-m Magellan Clay
    Telescope located at Las Campanas Observatory, Chile; various
    telescopes at Lick Observatory; the 1.3-m PAIRITEL on Mt. Hopkins;
    the SMARTS Consortium 1.3-m telescope located at Cerro Tololo
    Inter-American Observatory (CTIO), Chile; the 3.6-m
    Canada-France-Hawaii Telescope (CFHT), which is operated by the
    National Research Council of Canada, the Institut National des
    Sciences de l'Univers of the Centre National de la Recherche
    Scientifique of France, and the University of Hawaii; and the
    W. M. Keck Observatory, which is operated as a scientific
    partnership among the California Institute of Technology, the
    University of California, and NASA, with generous financial
    support from the W. M. Keck Foundation.}}


\author{Nancy Elias-Rosa\altaffilmark{2},
  Schuyler D.~Van Dyk\altaffilmark{2},
  Weidong Li\altaffilmark{3},
  Adam A.~Miller\altaffilmark{3},
  Jeffrey M.~Silverman\altaffilmark{3},
  Mohan Ganeshalingam\altaffilmark{3},
  Andrew F.~Boden\altaffilmark{4},
  Mansi M.~Kasliwal\altaffilmark{4},
  J\'ozsef Vink\'o\altaffilmark{5,6},
  Jean-Charles Cuillandre\altaffilmark{7},
  Alexei V. Filippenko\altaffilmark{3},
  Thea N. Steele\altaffilmark{3},
  Joshua S. Bloom\altaffilmark{3},
  Christopher V. Griffith\altaffilmark{3},
  Io K. W. Kleiser\altaffilmark{3}, and
  Ryan J. Foley\altaffilmark{8,9}
}

\altaffiltext{2}{Spitzer Science Center, California Institute of
  Technology, 1200 E. California Blvd., Pasadena, CA 91125; email
  nelias@ipac.caltech.edu.}
\altaffiltext{3}{Department of Astronomy, University of California,
  Berkeley, CA 94720-3411.}
\altaffiltext{4}{Division of Physics, Math, and Astronomy, California
  Institute of Technology, Pasadena, CA 91125.}
\altaffiltext{5}{Department of Optics \& Quantum Electronics,
  University of Szeged, D\'om t\'er 9, Szeged H-6720, Hungary.}
\altaffiltext{6}{Department of Astronomy, University of Texas, Austin,
  TX 78712.}
\altaffiltext{7}{Canada-France-Hawaii Telescope Corporation, 65-1238
  Mamalahoa Hwy, Kamuela, HI 96743.}
\altaffiltext{8}{Harvard/Smithsonian Center for Astrophysics, 60
  Garden Street, Cambridge, MA 02138.}
\altaffiltext{9}{Clay Fellow.}

\begin{abstract}

We present early-time photometric and spectroscopic observations of
supernova (SN)~2009kr in NGC~1832. We find that its properties to date
support its classification as Type II-linear (SN~II-L), a relatively
rare subclass of core-collapse supernovae (SNe). We have also
identified a candidate for the SN progenitor star through comparison
of pre-explosion, archival images taken with WFPC2 onboard the 
{\sl Hubble Space Telescope\/} with SN images obtained using adaptive
optics (AO) plus NIRC2 on the 10-m Keck-II telescope. Although the
host galaxy's substantial distance ($\sim$26 Mpc) results in large
uncertainties in the relative astrometry, we find that if this
candidate is indeed the progenitor, it is a highly luminous ($M_V^0 =
-7.8$ mag) yellow supergiant with initial mass $\sim$18--24
M$_{\sun}$.  This would be the first time that a SN~II-L progenitor
has been directly identified. Its mass may be a bridge between the
upper initial mass limit for the more common Type II-plateau SNe (SNe
II-P) and the inferred initial mass estimate for one Type II-narrow SN
(SN IIn).

\end{abstract}


\section{Introduction}\label{introduction}

It is not yet exactly clear how to map massive stars of a given mass
range to a core-collapse supernova (CC~SN) subtype. We now know with a
growing degree of confidence that solitary stars in the range of $\sim
8$--$16\ {\rm M}_{\sun}$ inevitably explode as Type II-plateau
supernovae (SNe~II-P; e.g., \citealt{smartt09}), with much of their
hydrogen envelope still intact (see \citealt{filippenko97} for a
discussion of SN classification). We also have evidence that the Type
II-narrow (IIn) SN 2005gl had a luminous ($M_V=-10.3$ mag), very
massive (possibly with initial mass $M_{\rm ini} >50\ {\rm M}_{\sun}$)
progenitor star that exploded while in the luminous blue variable
phase \citep{galyam09}. This progenitor, the only example so far
identified for a SN~IIn, likely had a far smaller fraction of its
outer H layer remaining than do SNe~II-P. Various indirect clues may
indicate that at least some of the Type Ib/c SNe are connected to the
Wolf-Rayet phase, which is expected to occur for stars with $M_{\rm
  ini} >25$--$30\ {\rm M}_{\sun}$ (e.g., \citealt{crowther07}). That
leaves the Type II-linear (II-L) SNe with no directly known progenitor
star, as well as the range $M_{\rm ini} = 18$--$30 {\rm M}_{\sun}$
(exactly the range that \citealt{smartt09} dubbed ``the red supergiant
problem'') without a well-established endpoint. From the rates derived
with the Lick Observatory Supernova Search (LOSS) for a sample of
well-studied SNe (Li et al.\ 2010, in prep.), we know that the
majority of massive stars end their lives as SNe~II-P, while the
incidences of SNe~II-L and SNe~IIn are comparatively rare ($\sim$ 7\%
and $\sim$ 6\% of all CC~SNe, respectively).

Here we examine the case of SN~2009kr in NGC~1832. SN~2009kr was
discovered by \cite{itagaki09} on 2009 Nov. 6.73 (UT dates are used
throughout) and was spectroscopically classified as a SN~IIn
\citep{tendulkar09}, and then as a ``young type-II'' SN
\citep{steele09}. \cite{li09b} first identified a possible progenitor
star in archival {\sl HST\/} images from 2004, using as reference a
combined 160-s $r'$ image from CFHT+MegaCam\footnote[10]{A joint project
  of CFHT and CEA/DAPNIA.} on Nov. 21.49. Here we show that both
early-time photometric and spectroscopic observations strongly suggest
that this object is a SN~II-L, and we increase our confidence in the
progenitor identification via further comparison of the {\sl HST\/}
images with Keck-II/NIRC2 adaptive-optics (AO) data.


\section{The Early-Time Nature of SN~2009kr}\label{specph}

\subsection{Photometry}\label{specph_ph}

Optical $BVRI$ images of SN~2009kr were obtained with the Lick
Observatory 0.76-m Katzman Automatic Imaging Telescope (KAIT;
\citealt{filippenko01}) and the 1.0-m Nickel telescope, and
SMARTS+ANDICAM at CTIO.  They were all initially reduced using
standard procedures (see, e.g., \citealt{poznanski09}). Because of our
follow-up campaign on SN 2004gq \citep{modjaz07}, also in NGC~1832, we
had a well-calibrated photometric sequence in the host-galaxy field
and template images in all passbands for image subtraction. We used an
image-reduction pipeline \citep{ganesh10} to reduce all data and
calibrate the photometry to the standard Johnson $BV$ and Cousins $RI$
system.

Near-infrared (NIR) observations were obtained with the 1.3-m Peters
Automated Infrared Imaging Telescope (PAIRITEL; \citealt{bloom06}) and
reduced using standard procedures. We used archival 2MASS images of
NGC 1832 as templates to subtract from the SN images using
HOTPANTS\footnote[11]{http://www.astro.washington.edu/users/becker/hotpants.html.}
and calibrated the $JHK_s$ photometry against 2MASS stars in the
field.

The $BVRIJHK_{\rm s}$ light curves are shown in
Figure~\ref{fig_phspec}(a), relative to $B$-band maximum ($15.95 \pm
0.02$ mag on Nov. $13 \pm 1$, or JD 2,455,149 $\pm 1$). For
comparison, we also show the light curves of the SNe~II-L~1979C
\citep{balinskaia80,devau81,barbon82b}, 1980K \citep{buta82,dwek83},
1990K \citep{cappellaro95}, and 2001cy (unpublished KAIT data); the
SNe~II-P~1999em \citep{hamuy01,leonard02} and 1992H
\citep{clocchiatti96}; and,the possible SN~II-L~2000dc
\citep{poznanski09}.  SN~2009kr does not follow the SN~II-P plateau;
instead, it more closely resembles the SN~II-L linear decline (see
\citealt{barbon79}). A steeper decline is present in $VRI$ after
$\sim 65$ days in the SN~2009kr light curves; however, this decline
was also observed for SN~1980K in $V$. From our adopted distance and
total extinction (see \S \ref{specph_spec}), SN~2009kr reached
$M(B_{\rm max}) = -16.48\pm0.30$ mag, typical of SNe~II-L
\citep{young89}. We thus conclude that SN 2009kr displays the
photometric behavior of a SN~II-L.

\subsection{Spectroscopy}\label{specph_spec}

Low-resolution spectra of SN~2009kr were obtained on 2009 Nov. 10 and
25, Dec. 9 and 18, and 2010 Jan. 20 with the Lick 3-m Shane
telescope+Kast spectrograph \citep{kastref}; on Jan. 8 with the Clay
telescope+MagE \citep{marshall08}; on Feb. 15 with the 10-m Keck-II
telescope+DEIMOS; and on Mar. 9 with the 10-m Keck-I telescope+LRIS. 
The spectra were reduced and calibrated using standard
procedures (e.g., \citealt{matheson00}). The observing conditions were
not photometric, resulting in uncertain absolute-flux calibrations. We
observed with the slit placed at the parallactic angle \citep{fil82}.

\begin{figure*}
\centering
\includegraphics[height=3.5truein,width=2.7truein,angle=-90]{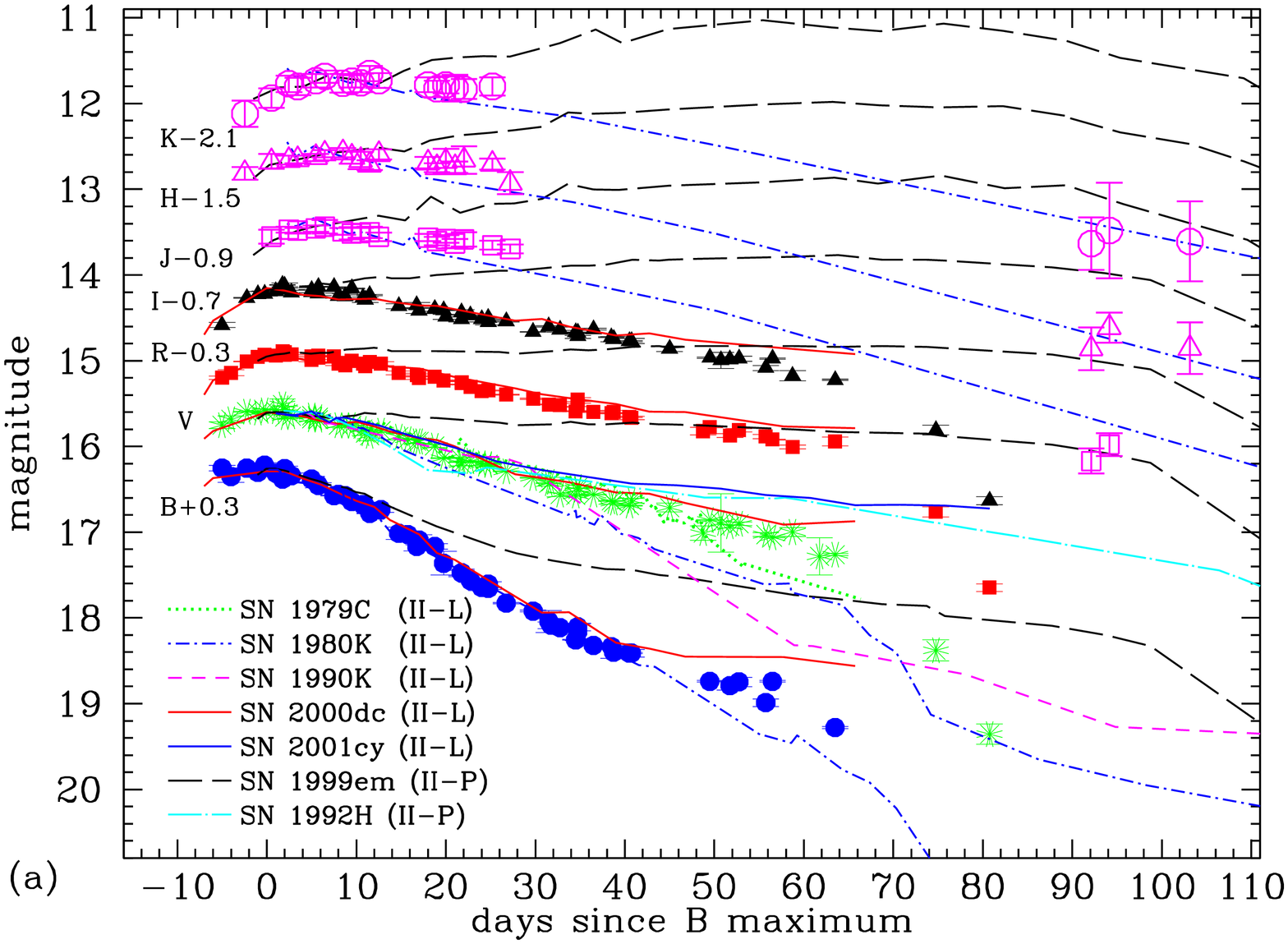}
\includegraphics[height=3.6truein,width=2.7truein,angle=-90]{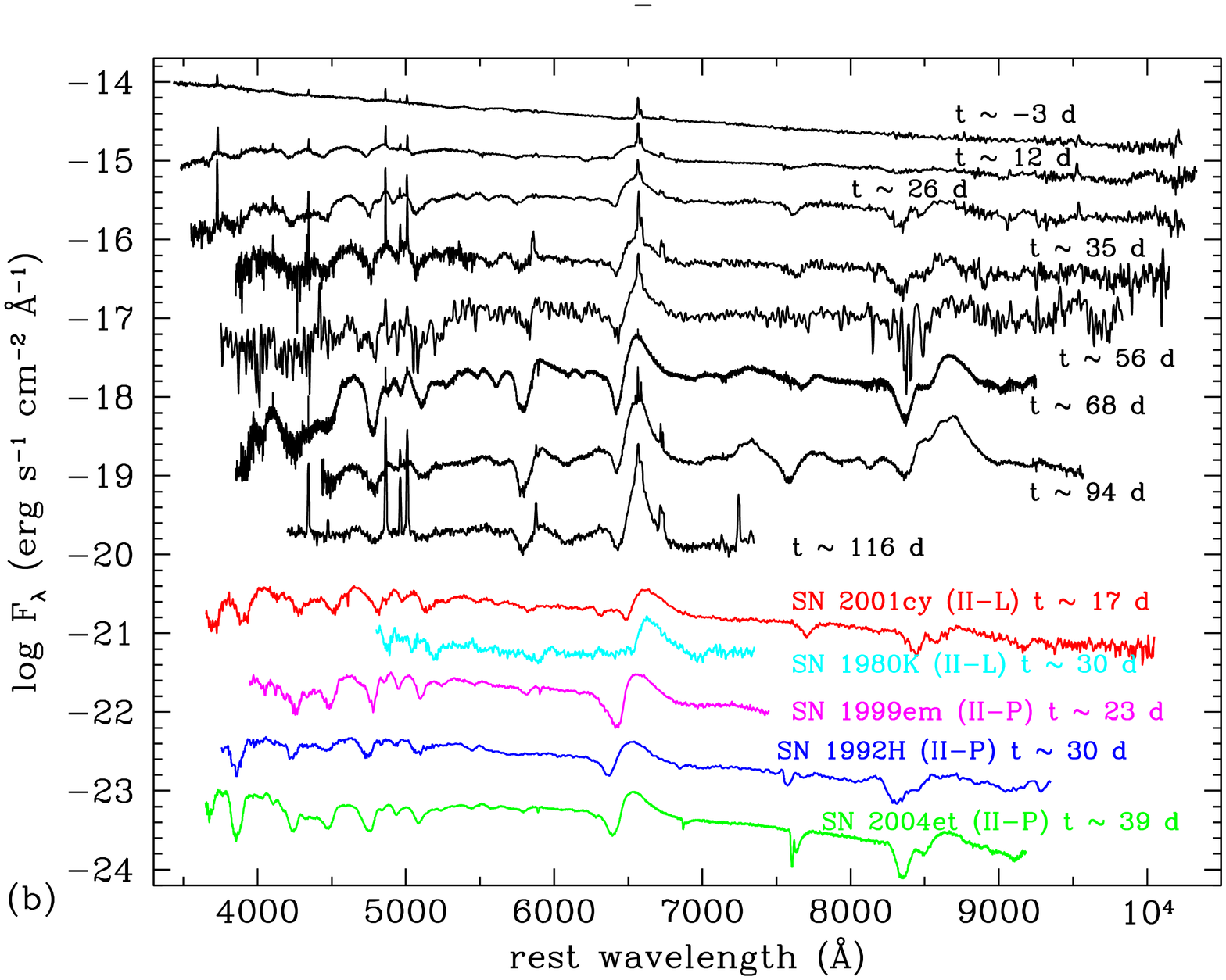}
\caption{{\it (a)} SN~2009kr optical and NIR light curves, together
  with those of other SNe~II-L and SNe~II-P. The comparison light
  curves are adjusted in time and magnitude to match those of SN
  2009kr.  {\it (b)} SN~2009kr spectral sequence, along with
  comparison spectra of other SNe~II-L and SNe~II-P.  All spectra have
  been corrected for their host-galaxy recession velocities and for
  reddening (see text). Ages are relative to $B$ maximum light.
  The continuum of the $t \approx 116$ day spectrum of SN 2009kr 
  appears to be significantly contaminated by light from the nearby
  star cluster.}
\label{fig_phspec}
\end{figure*}

We show the rest-frame spectral sequence in
Figure~\ref{fig_phspec}(b); for comparison, we also display spectra of
the SN~II-L~1980K \citep{barbon82a}; the SNe~II-P~1999em
\citep{leonard02}, 1992H \citep{clocchiatti96} and 2004et
\citep{sahu06}; and the SN~II~2001cy at similar
epochs. \cite{poznanski09} rejected SN~2001cy from their SN~II-P
sample, implying that it likely may be a SN~II-L. Indeed, the
SN~2001cy light curve is very similar (but of inferior quality and
coverage) to that of SN 2000dc shown in Figure~\ref{fig_phspec}(a).
The following corrections for $E(B-V)_{\rm tot}$ have been made:
SN~1999em (0.10 mag, \citealt{leonard02}), SN~1992H (0.09 mag,
\citealt{clocchiatti96}), SN~2004et (0.41 mag, \citealt{maguire09}),
SN~1980K (0.40 mag, \citealt{barbon82a}), and SN~2009kr (0.08 mag; see
below). SN~2001cy has been corrected only for Galactic reddening,
$E(B-V)_{\rm Gal} = 0.21$ mag \citep{schlegel98}.

The spectral sequence for SN~2009kr is typical of many SNe~II: the
earliest spectra exhibit a blue continuum, with relatively weak
spectral features, followed by the onset of increased Balmer-line
emission, together with the emergence of P-Cygni-like
features. However, clearly SN~2009kr differs from the canonical
SN~1999em or the peculiar SNe~1992H and 2004et; in particular, the
SN~2009kr P-Cygni H$\alpha$ profile is dominated by the broad emission
component. Several narrow emission lines appear superposed on the
spectra, but these most likely originate from a neighboring H\,{\sc ii}
region, seen $\sim 1$--2$\arcsec$ northeast of the SN in an H$\alpha$ 
image of the host galaxy (unpublished ESO archival data; the region's
ionizing cluster can also be seen in the {\sl HST\/} images). The
SN~2009kr spectra more closely resemble those of SN~1980K or
SN~2001cy, which also exhibit relatively weak H$\alpha$
absorption. The optical photometric decline rates [$\ga 1$ mag (100
  day)$^{-1}$] of SNe~1980K, 2001cy, and 2009kr suggest that they are
all SNe~II-L (see \S~\ref{specph_ph}). Since spectroscopic features of
SNe~II appear to be correlated with their photometric behavior at
early times \citep{schlegel96,filippenko97}, the similarities in the
spectra of these three SNe are not unexpected.

We also obtained a high-resolution optical spectrum of SN~2009kr on
2009 Nov. 25.46 with CFHT+ESPaDOnS, with 4 exposures of 833~s each,
from which we confirm that the narrow H$\alpha$ emission likely
originates from the H\,{\sc ii} region.  Moreover, we also used this
spectrum to estimate the reddening toward SN 2009kr via measurement of
the Na\,{\sc i}~D equivalent width (EW) at the host-galaxy redshift
($z=0.006$). We found EW(Na\,{\sc i}~D1 $\lambda$5890) = $0.044 \pm
0.003$ \AA\ and EW(Na\,{\sc i}~D2 $\lambda$5896) = $0.032 \pm 0.004$
\AA.  Using the relation between extinction and EW(Na\,{\sc i} D) from
Elias-Rosa et al.\ (2010, in prep.), and assuming the
\citet{cardelli89} reddening law with updated wavelengths and a
Galactic foreground $E(B-V)$ = 0.07 mag (Schlegel et al.\ 1998), we
derive ${E(B-V)}_{\rm tot} = 0.08 \pm 0.01$ mag (${E[V-I]}_{\rm tot} =
0.11 \pm 0.01$ mag), which we adopt for the SN.  This relatively low
extinction is consistent with both the SN color comparison
(\S~\ref{specph_ph}) and the overall blue continua seen in the early-time
SN~2009kr spectra (Fig. \ref{fig_phspec}[b]).


\section{Identification of the Progenitor Candidate}\label{identification}

Pairs of {\sl HST\/} images of NGC 1832 were obtained with WFPC2 in
bands F555W ($\sim V$; 460~s total) and F814W ($\sim I$; 700~s total)
on 2008 Jan. 11 (program GO-10877, PI: W.~Li), as SN~2004gq follow-up
observations.  The SN~2009kr site is located on the WF3 chip
($0{\farcs}1$ pixel$^{-1}$). Cosmic-ray hits were rejected, and a
$1600 \times 1600$ pixel mosaic of all four WFPC2 chips was
constructed using the STSDAS package routines {\it crrej} and 
  {\it wmosaic} within IRAF\footnote[12]{IRAF (Image Reduction and Analysis
  Facility) is distributed by the National Optical Astronomy
  Observatories, which are operated by the AURA, Inc., under
  cooperative agreement with the National Science
  Foundation (NSF).}. \cite{li09b} were able to isolate the SN location in
the WFPC2 images to $0.43$ pixel ($0\farcs043$) through comparison
with a ground-based, post-explosion CFHT+MegaCam image.

We were then able to better confirm the identification of this
candidate through $K_{\rm p}$-band NIRC2 ``wide'' camera ($0{\farcs}04$
pixel$^{-1}$, $40\arcsec \times 40\arcsec$ field of view) images
obtained on 2009 Nov. 28 with Keck-II+AO.  Each of the 10-s frames was
sky subtracted using the median of the dithered exposures, and then
``shifted-and-added'' using IRAF. They were also corrected for
distortion\footnote[13]{http://www2.keck.hawaii.edu/inst/nirc2/forReDoc/post\_observing/dewarp/.}.

\begin{figure*}
\epsscale{1.0} 
\plotone{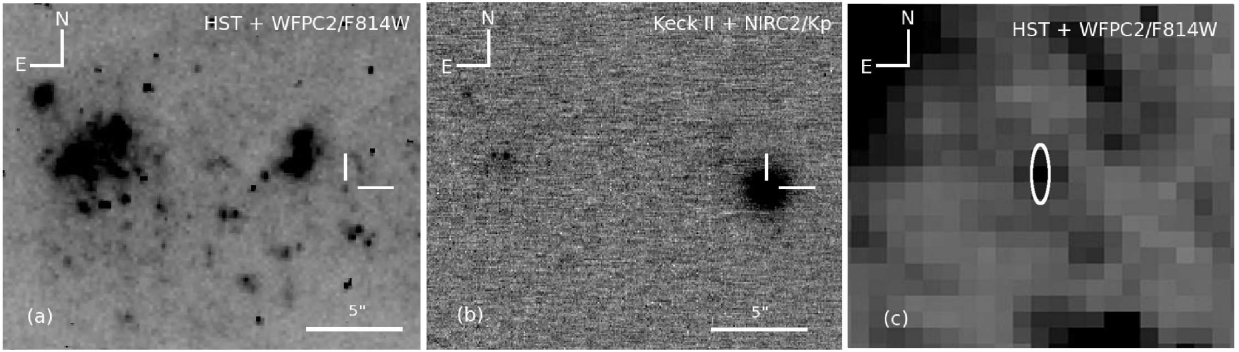}
\caption{(a) Subsections of the pre-explosion {\sl HST\/}+WFPC2/F814W
  images of NGC~1832, and (b) the post-explosion AO $K_{\rm p}$ image of
  SN~2009kr with Keck II+NIRC2. The approximate positions of the
  candidate progenitor and the SN are indicated by tick~marks. (c)
  $5\arcsec \times 5\arcsec$ detail of the {\sl HST\/} image. The
  3$\sigma$ positional uncertainty ellipse is $0{\farcs}03 \times
  0{\farcs}09$ in radius (see Table~\ref{table_error}).}
\label{fig_progenitor}
\end{figure*}

We achieved high-precision relative astrometry by geometrically
transforming the pre-explosion images to match the post-explosion
ones. We first
``drizzled''\footnote[14]{http://www.stsci.edu/hst/wfpc2/analysis/drizzle.html.}
the pre-explosion images for each band to the higher NIRC2+AO
resolution. Using 5--7 point-like sources in common between the two
datasets and the IRAF tasks {\it geomap\/} and {\it geotran\/}, we
carried out a geometrical transformation between the two sets of
images. The positions (and their uncertainties) of the SN and the
progenitor candidate are derived by averaging the measurements from
two centroiding methods, the task {\it daofind\/} within IRAF/DAOPHOT
and {\it imexamine} within IRAF.  The differences between the SN and
the progenitor candidate positions, compared with the total estimated
astrometric uncertainty, are given in Table~\ref{table_error}. As an
additional check, we transformed the NIRC2+AO image relative to only
the WF3 chip image in F814W at its native resolution ($0{\farcs}1$
pixel$^{-1}$) and performed the registration. In this case, the
positional difference is 19 mas, within $\sim 1\sigma$ of the
uncertainty in the astrometric solution. Note that no other source was
located within a $5\sigma$ radius from the progenitor candidate
position identified by Li et al.\ (2009); the H\,{\sc ii} region to the
northeast is located at $> 25\sigma$ (see Fig.
\ref{fig_progenitor}).


\section{The Nature of the Progenitor}\label{discussion}

We also measured photometry of the {\sl HST\/} images using
HSTphot\footnote[15]{HSTphot is a stellar photometry package designed
  for use with WFPC2 images. We used v1.1.7b, updated 2008 July 19.}
(\citealt{dolphin00}). The output from this package automatically
includes the transformation from F555W and F814W to $V$ and $I$.

Adopting a distance modulus to NGC~1832 derived from the recession
velocity corrected for Virgo infall ($32.09 \pm 0.30$ mag; this
uncertainty arises from a possible 250 km s$^{-1}$ peculiar
velocity\footnote[16]{From NED, http://nedwww.ipac.caltech.edu/.}) and
the assumed SN extinction, we find that the absolute magnitudes of the
progenitor candidate are $M_V^0=-7.80 \pm 0.33$ and $M_I^0=-8.75 \pm
0.32$, entirely consistent with a highly luminous supergiant. The
intrinsic color, $(V-I)_0=0.95 \pm 0.21$ mag, is significantly more
``yellow'' than the colors of normal red supergiants (RSGs; e.g.,
\citealt{drilling00}).

We can directly determine the metallicity in the SN environment from
the CFHT high-resolution spectrum, by measuring the H$\alpha$ and
[N~II] $\lambda$6584 line intensities and applying the
\citep{pettini04} cubic-fit relation between the ratio of these two
lines and the oxygen abundance; no correction is made for the low
extinction. Doing so, we find that 12 + log(O/H) = 8.67. Given that
the solar value is 8.69 $\pm$ 0.05 \citep{asplund09}, we consider it
most likely that this environment has roughly solar metallicity.

\begin{figure*}
\epsscale{1.0}
\centering
\includegraphics[height=5truein,width=3.8truein,angle=-90]{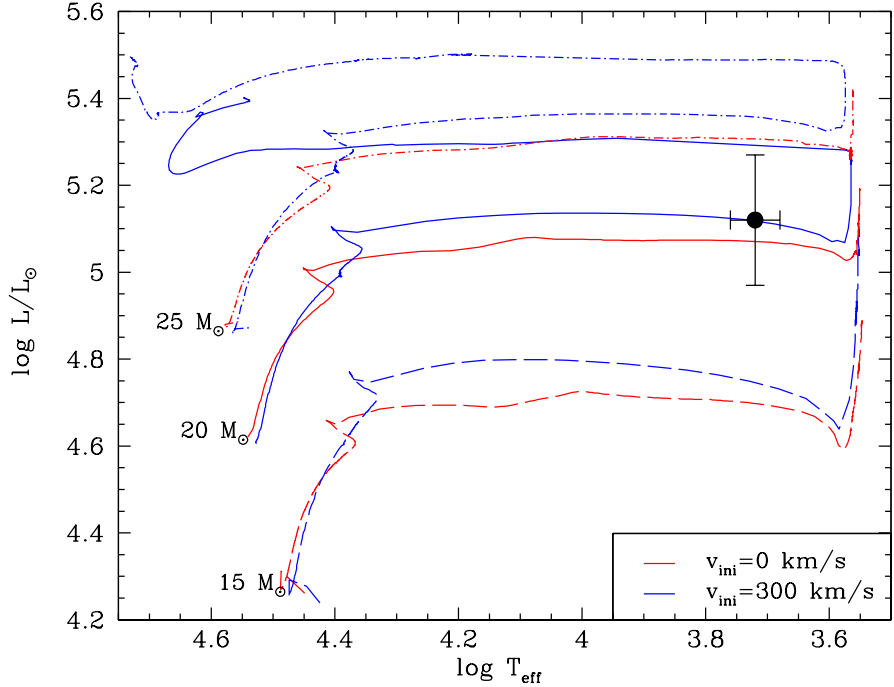}
\caption{
An HR diagram showing the $L_{\rm bol}$ and $T_{\rm eff}$ for the
candidate progenitor of SN~2009kr ({\it filled circle}). Model stellar
evolutionary tracks for solar metallicity (Hirschi et al.\ 2004) are
also shown for a rotation of $v_{\rm ini} = 0$ \kms ({\it dot-dashed},
{\it solid}, and {\it dashed red lines}) and $v_{\rm ini} = 300$
\kms\ ({\it dot-dashed}, {\it solid}, and {\it dashed blue lines}).
}
\label{fig_hrd}
\end{figure*}

The star's color corresponds to an effective temperature $T_{\rm eff}
= 5300 \pm 500$ K and a $V$ bolometric correction in the range $-0.47$
to $-0.13$ mag, for an assumed surface gravity $\log g = +0.5$ (Kurucz
Atlas 9 models, CD-ROMs 13, 18).  This results in $L_{\rm bol} =
10^{(5.12 \pm 0.15)}\ {\rm L}_{\sun}$ (assuming $M_{\rm bol}({\sun})$=
4.74 mag). In Figure~\ref{fig_hrd} we show a Hertzsprung-Russell (HR)
diagram including the progenitor candidate. In addition, we illustrate
model evolutionary tracks (\citealt{hirschi04}) for stars with $M_{\rm
  ini}=15$, 20, and 25~M$_{\sun}$, with rotation ($v_{\rm ini} = 300$
\kms) and without rotation.

The location of the progenitor candidate in the HR diagram is clearly
not consistent with the 15~M$_{\sun}$ tracks, an initial mass which
lies within the range for SN~II-P progenitors
(\citealt{smartt09}). Unfortunately, the mass bins for these tracks
are large (5~M$_{\sun}$ increments).  However, taking into account the
uncertainty in the progenitor candidate's luminosity, interpolating by
eye between the tracks implies that $M_{\rm ini} \approx
18$--$24\ {\rm M}_{\sun}$.  We note that this is consistent with the
upper limit of $<20\ {\rm M}_{\sun}$ on the SN 1980K progenitor
(\citealt{smartt09}, using more current theoretical tracks) and the
lower limit of $\gtrsim 17$--18 $M_{\sun}$ on the SN II-L 1979C
progenitor (\citealt{vandyk99}).

That such a massive progenitor would be somewhat bluer than would be
expected for the normal RSG progenitors of SNe~II-P is also consistent
with the star being in a post-RSG phase. Such an expectation follows from
the theoretical models by Hirschi et al.\ (2004) for
rotating stars at higher masses; the rotating
models are more luminous and evolve to the RSG phase before the
ignition of He burning, which results in higher mass-loss rates, the
loss of most of the H envelope before the termination of He
burning, and evolution toward the blue before the terminal points.
From Figure~\ref{fig_hrd} one can see that rotation makes
little difference in the evolution of a 15~M$_{\sun}$ star; both the
rotating and nonrotating tracks terminate as a RSG. However, for a 
20~M$_{\sun}$ star, rotation has a profound evolutionary effect. 
Though we cannot tell from these models what occurs for stars in the
15--20~M$_{\sun}$ range, it is reasonable to assume that rotation
begins to affect those models nearing 20~M$_{\sun}$. Thus, we might
expect the progenitor candidate, based on our mass estimate above, to
have evolved toward the blue before explosion.

Although more luminous and hotter than the SN 2009kr progenitor
candidate, similar behavior has occurred for the post-RSG star
IRC+10420, a mass-losing hypergiant that is transiting the so-called
``yellow void'' (\citealt{humphreys02}). Likely more analogous are the
``anomalous'' yellow supergiants in the Magellanic Clouds, which are
in the mass range 15--20~M$_{\sun}$ and show evidence for post-RSG
evolution (\citealt{humphreys91}), albeit at subsolar metallicity.

Of course, there are caveats; as we found for the peculiar
SN~II-P~2008cn (\citealt{eliasrosa09}), a yellow progenitor color
could result from evolution in an interacting binary. Additionally,
although the difference between the SN and progenitor candidate
positions are practically within the total uncertainties, these are
still large, and, given the host galaxy's distance (a single NIRC2
pixel corresponds to $\sim 5$ pc), it is possible that we have not
identified the SN progenitor at all, but rather a compact star
cluster.

\begin{figure*}
\plotone{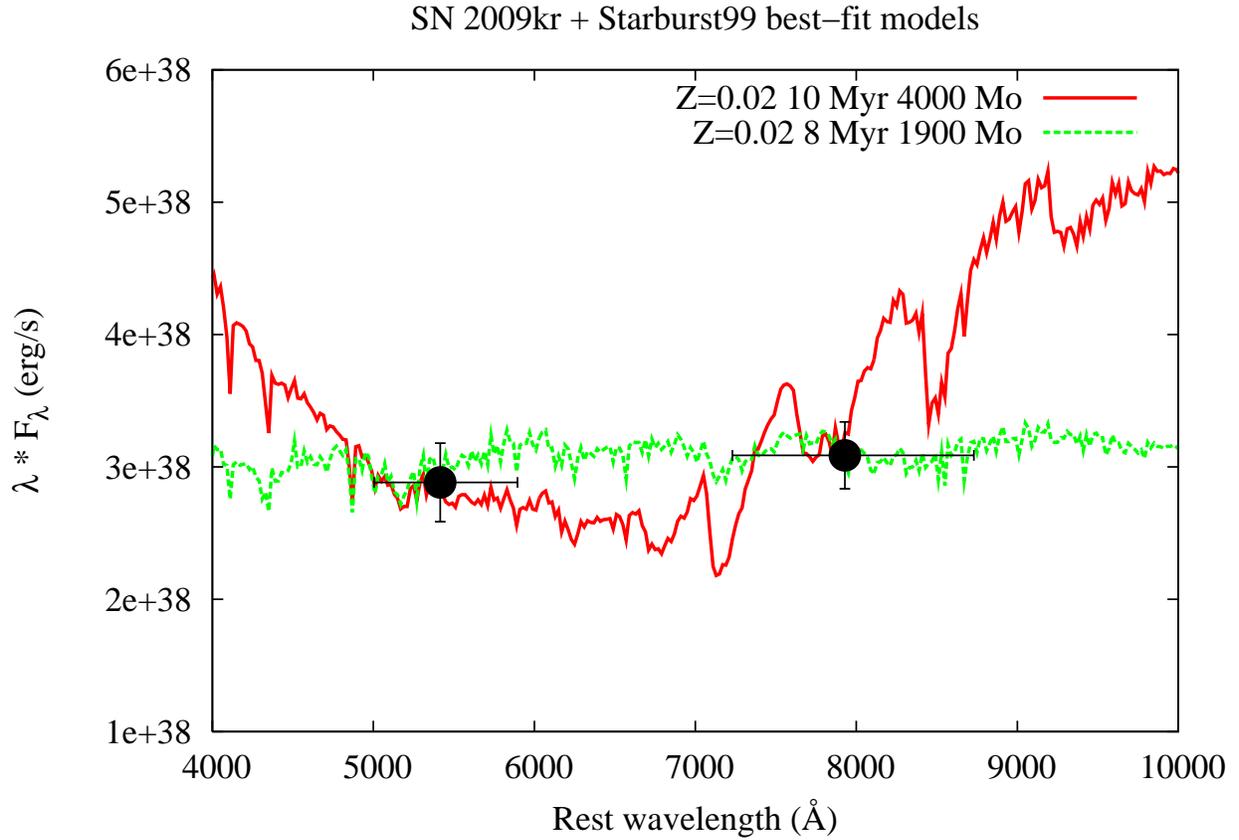}
\caption{Comparison of the progenitor candidate ({\it filled circles})
  with model spectral energy distributions for star clusters
  (Starburst99; \citealt{vazquez05}; {\it lines}) for solar
  metallicity. The model parameters (age and total mass) are indicated
  in the legend. }
\label{fig_sed}
\end{figure*}

If so, adopting the SN extinction, we can fit the $V$ and $I$ fluxes
for the progenitor candidate with the Starburst99
code\footnote[17]{http://www.stsci.edu/science/starburst99/,
  \cite{vazquez05}.} model spectral energy distributions and find an
excellent fit with a 10~Myr track (Fig.~\ref{fig_sed}). If the SN
progenitor were a member of this cluster, this age estimate would
still be consistent with the lifespan of a $\sim$20 M$_{\sun}$
star. The 8~Myr model also provides a good fit. However, the lowest
possible cluster luminosity that exceeds the luminosity of the most
massive 8 Myr-old star is $M_V \approx -8.1$ mag, and for 10 Myr it is
$M_V \approx -7.6$ mag \citep{cervi04}. Both of these are roughly
consistent with the candidate's luminosity.  Furthermore, we would
ordinarily expect compact star clusters to have $M_V < -8.6$ mag
(\citealt{bastian05,crockett08}). Additional dust obscuration, beyond
our assumed extinction, would imply that a putative cluster could be
more luminous, but it would also be bluer and younger.

A similar study has been done contemporaneously by \cite{fraser09},
identifying the same candidate progenitor. However, \cite{fraser09}
suggest that the SN is a spectrally peculiar SN II-P, with only
$r$-band photometry reported. Differences also exist in the
metallicity estimate and theoretical models employed, leading to
somewhat different estimates for $M_{\rm ini}$
($15^{+5}_{-4}\ {\rm M}_{\sun}$ by \citealt{fraser09}).

\section{Conclusions}\label{conclusions}

Based on $\sim3$ months of follow-up observations from discovery, we
conclude that SN~2009kr is a SN~II-L, a relatively rare subclass of
CC~SNe.  We identified an object in pre-explosion {\sl HST\/} images
which agrees astrometrically with the SN~2009kr position, using NIR SN
images obtained with NIRC2+AO on Keck.  From the SN distance and
extinction we find that the object is consistent with a highly
luminous supergiant star. Placing the inferred $L_{\rm bol}$ and
$T_{\rm eff}$ of the progenitor candidate on an HR diagram, we infer
$M_{\rm ini} \approx 18$--$24\ {\rm M}_{\sun}$.  Its yellow color
implies that the star may have exploded in a post-RSG evolutionary
state, which is predicted by our assumed theoretical stellar models.
If this star is the SN~2009kr progenitor, this would be {\it the
  first-ever direct identification of a SN II-L progenitor}. The
star's mass estimate also may be a link between the upper mass range
for SNe~II-P (\citealt{smartt09}) and the estimated progenitor mass
for SN~IIn~2005gl (\citealt{galyam09}).

Ultimately, the definitive indication that we have identified the SN
2009kr progenitor is through very late-time imaging, such as with 
{\sl HST\/} (\citealt{maund09}).  However, as a SN~II-L, akin to the
long-lasting SN 1979C (\citealt{mili09}), we may have to wait at least
a decade to image the SN site using the {\sl James Webb Space Telescope}.


\acknowledgments

We thank C. Blake, S. B. Cenko, B. Cobb, E. Falco, M. Kandrashoff,
J. Kong, M. Modjaz, D. Starr, and T. Yuan for their
assistance. J.V. received support from Hungarian OTKA Grant K76816,
NSF Grant AST-0707769, and Texas Advanced Research Project grant
ARP-0094. A.V.F.'s group and KAIT are supported by NSF grant
AST-0908886, the Sylvia \& Jim Katzman Foundation, the TABASGO
Foundation, and NASA through grants AR-11248 and GO-10877 from STScI.
PAIRITEL is operated by SAO with support from the Harvard University
Milton Fund, UC Berkeley, University of Virginia, and NASA/{\it Swift}
grant NNX09AQ66G. J.S.B. and his group are partially funded by a DOE
SciDAC grant.

{\it Facilities:} \facility{HST (WFPC2)}; \facility{Lick: KAIT, Nickel, Shane}; 
 \facility{CFHT: MegaCam}; \facility{Keck-I: LRIS}; \facility{Keck-II: NIRC2+AO, DEIMOS}; 
  \facility{CTIO: SMARTS};  \facility{Magellan: MagE}


\clearpage

\clearpage

\begin{deluxetable}{lcc}
\tablewidth{0pt}
\tablecaption{SN and Progenitor Candidate Position Comparison\label{table_error}}
\tablehead{
\colhead{}           & \colhead{F555W ($\alpha/\delta$)}      &
\colhead{F814W ($\alpha/\delta$)}
}
\startdata
Uncertainty in the progenitor position (mas) & 0/2 & 0/24\\
Uncertainty in the SN position (mas) & 8/13 & 8/13\\
Geometric transformation (mas) & 4/15 & 5/13\\
Total uncertainty (mas) & 9/20 & 9/30\\
\tableline
Difference in position (mas) & 13/4 & 7/22\\
\enddata
\tablecomments{Uncertainties ($1\sigma$) in the SN and progenitor
  candidate right ascension ($\alpha$) and declination ($\delta$)
  positions for each band, in milliarcsec (mas), were estimated as the
  standard deviation of the average measurements. Uncertainties in the
  geometric transformation were derived from the differences in the
  fiducial star positions before and after the transformation. The
  total uncertainty is the quadrature sum of all uncertainties. The
  last row lists the residual difference between the SN and progenitor
  position after the geometric transformation. See text.}
\end{deluxetable}

\end{document}